\documentclass[aps,pra,twocolumn,groupedaddress,showpacs]{revtex4}
\usepackage{amsfonts}
\usepackage{amssymb}
\usepackage{bm}
\usepackage{graphicx}%
\usepackage{hyperref}
\usepackage{amsmath}
\newcommand{\ket}[1]{\ensuremath{\left|{#1}\right\rangle}}

\newcommand{\bsy}[1]{\ensuremath{\boldsymbol{#1}}}
\newcommand{\brm}[1]{\ensuremath{\mathbf{#1}}}

\begin{document}

\title{Optical Bell-state analysis in the coincidence basis}


\author{S. P. Walborn}
\email[]{swalborn@fisica.ufmg.br}

\affiliation{Universidade Federal de Minas Gerais, Caixa Postal 702, 
Belo Horizonte, MG
30123-970, Brazil}
\author{W. A. T. Nogueira}
\affiliation{Universidade Federal de Minas Gerais, Caixa Postal 702, 
Belo Horizonte, MG
30123-970, Brazil}
\author{S. P\'adua}
\affiliation{Universidade Federal de Minas Gerais, Caixa Postal 702, 
Belo Horizonte, MG
30123-970, Brazil}
\author{C. H. Monken}
\affiliation{Universidade Federal de Minas Gerais, Caixa Postal 702, 
Belo Horizonte, MG
30123-970, Brazil}

\date{\today}

\begin{abstract}
Many quantum information protocols require a Bell-state measurement of entangled systems.  Most optical Bell-state measurements utilize two-photon   interference at a beam splitter.  By creating polarization-entangled 
photons with spontaneous parametric down-conversion using a 
first-order Hermite-Gaussian pump beam, we invert the usual interference behavior and perform an incomplete Bell-state measurement  in 
the coincidence basis.  We discuss the possibility of a complete Bell-state measurement in the coincidence basis using hyperentangled states  [Phys. Rev. A, \textbf{58}, R2623 (1998)].
 \end{abstract}
\pacs{03.67.-a, 03.67.Hk, 42.50.-p}
\maketitle
Many quantum information schemes require entangled Bell-states as a 
resource.  Furthermore, protocols such as dense coding 
\cite{bennett92,mattle96}, quantum teleportation 
\cite{bennett93,bouwmeester97,boschi98} and entanglement swapping 
\cite{bennett93,pan98,jennewein02} require a Bell-state measurement (BSM), that 
is, distinguishing between the four Bell-states: 
\begin{align}
& \ket{\psi^{\pm}}=\frac{1}{\sqrt{2}}\left(\ket{h}_{1}\ket{v}_{2} \pm 
\ket{v}_{1} \ket{h}_{2} \right), \\
& \ket{\phi^{\pm}}=\frac{1}{\sqrt{2}}\left(\ket{h}_{1}\ket{h}_{2} \pm 
\ket{v}_{1} \ket{v}_{2} \right), \label{eq:bellstates}
\end{align}  
where $\ket{\psi^{-}}$ is the antisymmetric singlet state and  
$\ket{\psi^{+}}$,$\ket{\phi^{\pm}}$ are the symmetric triplet states.
In this paper we use polarization Bell-states, so $h$ and $v$ stand 
for horizontal and vertical polarization and $1$ and $2$ represent 
different spatial modes.  These states are easily generated using 
spontaneous parametric down-conversion \cite{kwiat95,kwiat99}.
However, difficulty arises when one wants to distinguish between the 
four Bell-states.  In fact, it has been proven that a complete 
Bell-state measurement (discriminating between all four states with 100\% efficiency) using only linear optics is impossible
\cite{vaidman99,lutkenhaus99,calsamiglia01,ghosh01}.  A complete  BSM was realized in ref. \cite{boschi98}, but the entangled 
systems were two degrees of freedom of the same photon.  Using 
nonlinear optical processes, one can discriminate among the four states with low efficiency \cite{kim00a}.  It is possible to perform an incomplete BSM of polarization-entangled photons using two-photon interference and polarizing beam splitters \cite{mattle96,braunstein95}.    Another possible scheme is using photon absorption in 
properly prepared atoms \cite{scully99}.  Using hyperentangled 
states,  a complete Bell-state measurement is possible 
\cite{kwiat98a}, though this scheme, along with those used in 
\cite{mattle96,michler96} requires detectors sensitive to photon 
number.   
For momentum entanglement, there are several methods that allow one 
to  distinguish two 
\cite{weinfurter94,michler96} of the four 
states and yet another that discriminates between all four with 25\% efficiency \cite{weinfurter94}.
\par 
Expanding on previous results 
\cite{mattle96,kwiat98a}, we wish to report on methods for optical Bell-state measurement  
in the coincidence basis, that is, each photon triggering a different detector.  We use polarization-entanglement due to the ease with which one can generate and manipulate polarization-entangled photon pairs.   
\section{Incomplete Bell-state analysis in the coincidence basis}
\label{sec:ibsa}
\begin{figure}
 \includegraphics[width=8cm]{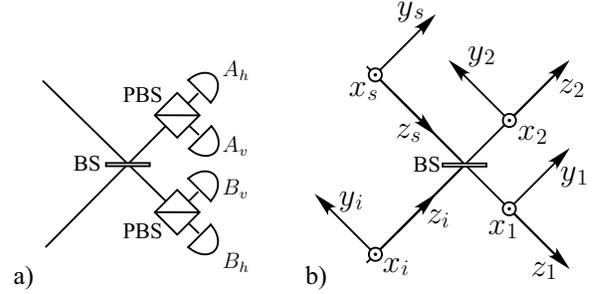}%
 \caption{a)  Incomplete Bell-state analyzer.  PBS are 
polarizing beam splitters at $0^{\circ}$.  b)  Two-photon interference at a beam splitter 
(BS).  Photons $s$ and $i$ pass through the BS into modes $1$ and $2$.}
\label{fig:BS-BSA}
\end{figure} 

Most optical BSM's \cite{mattle96,braunstein95} of 
polarization-entangled photons rely on Hong-Ou-Mandel type two-photon 
interference at a 50-50 beam splitter \cite{hom87}.   
Consider the Bell-state analyzer (BSA) shown in fig. \ref{fig:BS-BSA}a,  such as the 
one used in the experimental demonstration of dense coding 
\cite{mattle96}. This BSA is capable of separating the four Bell states into 3 classes, 
resulting in a ``trit" ($\approx 1.58$ bits) of transmitted 
information.  A 50-50 
beam splitter is used to separate $\ket{\psi^{-}}$ from 
$\ket{\psi^{+}}$, $\ket{\phi^{\pm}}$: bosonic symmetry requires that photons in $\ket{\psi^{-}}$ end up in different outputs while photons in $\ket{\psi^{+}}$, $\ket{\phi^{\pm}}$ end up in the same output \cite{zeilinger94}.  
 With the polarizing beam splitters (PBS) separating $h$ and $v$ 
polarizations, coincidences at $A_{h}B_{v}$ or $A_{v}B_{h}$ are the 
signature of $\ket{\psi^{-}}$.  The PBS separate $\ket{\psi^{+}}$ from 
$\ket{\phi^{\pm}}$:  coincidences at $A_{h}A_{v}$ or $B_{h}B_{v}$ are 
characteristic of $\ket{\psi^{+}}$.  For $\ket{\phi^{\pm}}$, we have two photons 
at $A_{h}A_{h}$, $A_{v}A_{v}$, $B_{h}B_{h}$  or $B_{v}B_{v}$.  Thus, detection of $\ket{\psi^{-}}$,  $\ket{\psi^{+}}$ and 
$\ket{\phi^{\pm}}$ requires detectors capable of distinguishing between one and two photons.  Such detectors are presently available, however they suffer from low efficiencies and/or high dark counts \cite{kwiat93,kim99,takeuchi99}.  As mentioned in refs. \cite{mattle96,michler96}, this problem can be partially solved by replacing each detector with two detectors and an additional 50-50 beam splitter.  This enables one to detect only half of the two-photon occurrences and increases the complexity of the detection system, since an eight detector system is necessary.            
  \par
We could avoid this requirement on the detectors if we could invert the  interference behavior:  photons in the triplet (singlet) states go to  different (the same) detectors and can then be further discriminated by the PBS.  This can be achieved by generating polarization-entangled photons using an 
antisymmetric pump beam, such as the first-order Hermite-Gaussian  
beam $\mathrm{HG}_{01}$, as we will show below.
 Then, pumping with $\mathrm{HG}_{01}$,
$\ket{\psi^{-}}$ results in 
two photons in either output port.  Since the two photons are 
orthogonally polarized, coincidences at detectors $A_{h}A_{v}$ or 
$B_{h}B_{v}$ are the signature of $\ket{\psi^{-}}$.   The states $\ket{\psi^{+}}$ and 
$\ket{\phi^{\pm}}$ give one photon in each output port.  Since the 
photon pairs of $\ket{\psi^{+}}$ are orthogonally polarized, $\ket{\psi^{+}}$ 
gives coincidence counts  at detectors  $A_{h}B_{v}$ or 
$B_{h}A_{v}$.   $\ket{\phi^{\pm}}$ results in coincidence counts at   
$A_{h}B_{h}$ or $A_{v}B_{v}$.       All  detector combinations identifying the three cases $\ket{\psi^{+}}$, $\ket{\psi^{-}}$ and $\ket{\phi^{\pm}}$ correspond to coincidences at different detectors.
Let us now discuss multimode interference and show how an antisymmetric pump beam inverts the interference behavior. 
A two-mode Bell-state, as given by  Eq. (\ref{eq:bellstates}), 
incident on a beam splitter as shown in fig. \ref{fig:BS-BSA}b,  
 must preserve its overall bosonic character. 
Thus if the polarization component of the state is symmetric 
(antisymmetric), then the spatial component  must also be symmetric 
(antisymmetric) \cite{zeilinger94}.  However, if we consider  
multimode fields, we must also take into account the transverse 
spatial properties of the two-photon state as additional degrees of 
freedom.  The combined transverse-spatial and polarization symmetry of the two-photon wave packet determines whether the fields will interfere constructively or destructively.     
Here we consider photon pairs created by spontaneous parametric 
down-conversion (SPDC) incident on opposite inputs of a beam splitter \cite{hom87,walborn02}.  A field reflected by the beam splitter undergoes a reflection in the horizontal ($y$) direction, while a transmitted field does not suffer any reflection, as illustrated in fig. \ref{fig:BS-BSA}b.  \par
 Using the standard theory of SPDC\cite{hong85,monken98a}, it can be shown that the multimode coincidence-detection amplitudes 
of the polarization Bell-states in the two outputs of a balanced HOM 
interferometer are given by \cite{walborn02}
\begin{widetext}
\begin{align}
\bsy{\Psi}^{\pm}(\brm{r}_{1},\brm{r}_{2})=&
\exp{\left\{\frac{iK}{2Z}\left[(x_{1}-x_{2})^{2}+(y_{1}-y_{2})^{2}\right]\right\}}
\nonumber \times \\
&\left [ 
\mathcal{W}\left(\frac{x_{1}+x_{2}}{2},\frac{y_{1}+y_{2}}{2},Z\right)
\mp 
\mathcal{W}\left(\frac{x_{1}+x_{2}}{2},\frac{-y_{1}-y_{2}}{2},Z\right) 
\right]  (\brm{h}\brm{v} \pm \brm{v}\brm{h}), \label{eq:psi}
\end{align}
\begin{align}
\bsy{\Phi}^{\pm}(\brm{r}_{1},\brm{r}_{2})=&
\exp{\left\{\frac{iK}{2Z}\left[(x_{1}-x_{2})^{2}+(y_{1}-y_{2})^{2}\right]\right\}}
\nonumber \times \\
&\left [ 
\mathcal{W}\left(\frac{x_{1}+x_{2}}{2},\frac{y_{1}+y_{2}}{2},Z\right)
- \mathcal{W}\left(\frac{x_{1}+x_{2}}{2},\frac{-y_{1}-y_{2}}{2},Z\right) 
\right]  (\brm{h}\brm{h} \pm \brm{v}\brm{v}), \label{eq:phi}
\end{align}
\end{widetext}
where $\bsy{\Psi}^{\pm}$ ($\bsy{\Phi}^{\pm}$) is the coincidence-detection amplitude for $\ket{\psi^{\pm}}$ ($\ket{\phi^{\pm}}$).  $\mathcal{W}(x,y,Z)$ 
is the transverse amplitude profile of the pump beam (propagated from 
$z=0$ to $z=Z$) transferred to the coincidence-detection amplitude 
\cite{monken98a}, $K$ is the magnitude of the pump beam wave vector, and 
$\brm{h}$ and $\brm{v}$ are unit polarization vectors in the $h$- and 
$v$-directions, respectively.  The vectors $\brm{r}_{1}=(x_{1},y_{1},z_{1})$ and $\brm{r}_{2}=(x_{2},y_{2},z_{2})$ correspond to the positions of the detectors and we have chosen experimental conditions such that $Z=z_{1}=z_{2}$.   Here we are working in the paraxial and monochromatic approximations, and have assumed that the beam splitter is $50$-$50$ and symmetric.  Furthermore, we disregard 
any entanglement between the momentum and polarization degrees of 
freedom \footnote{Entanglement between wave vector and polarization 
can be made negligible by using compensating crystals \cite{kwiat95} 
as well as small detection irises and narrow bandwidth interference 
filters.}.  The $(-y_{1}-y_{2})/2$ term in the $\mathcal{W}$ functions are a consequence of the reflection of both fields at the beam splitter.
\par
A quick look at (\ref{eq:psi}) and (\ref{eq:phi}) shows that when the 
pump beam profile is an even function of $y$, only 
$\ket{\psi^{-}}$ gives coincidence counts at the outputs of the BS (fig. \ref{fig:BS-BSA} a)).  If the pump beam profile is an 
odd function of $y$, then $\ket{\phi^{\pm}}$ and $\ket{\psi^{+}}$   
give coincidence counts, while $\ket{\psi^{-}}$ results in two 
photons in the same output of the beam splitter.  Thus,
using a pump beam profile that is an odd function of $y$ gives the desired interference effects for our BSM scheme mentioned above.   
\par
A well known class of beams with cartesian parity are the 
Hermite-Gaussian (HG) beams, 
\begin{align}
\mathrm{HG}_{mn}(x,y,z) = &  C_{mn}
H_{m}(x\sqrt{2}/w)H_{n}(y\sqrt{2}/w) \times \nonumber \\ & e^{(x^2+y^2)/w^2}  e^{-ik(x^2+y^2)/2R} e^{-i(m+n+1)\theta},
\end{align}
where $C_{mn}$ is a normalization constant. The $H_{n}(y)$ are the Hermite 
polynomials, which
are even or odd functions in the $y$-coordinate when the index $n$ is 
even
or odd, respectively. $w$ is the beam radius,
$R(z) = (z^2+z_{R}^2)/z$ and
$\theta(z) =\arctan(z/z_{R})$,
where $z_{R}$ is the Rayleigh range.
Using a pump beam in the $\mathrm{HG}_{01}$ mode, we can perform an incomplete Bell-state analysis in the coincidence basis.
\begin{figure}
 \includegraphics[width=7cm]{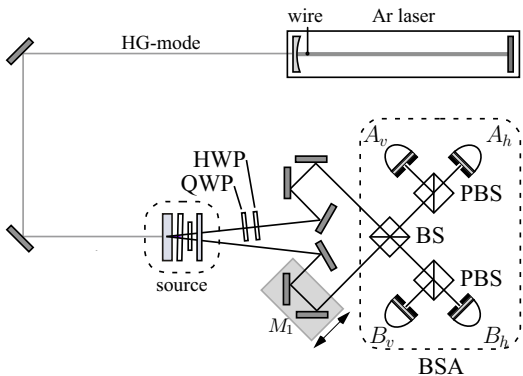}%
 \caption{ Experimental setup.  The source is described in the text.  BSA is the 
Bell-state analyzer consisting of a beam splitter (BS) and two 
polarizing beam splitters (PBS).}
\label{fig:setup}
 \end{figure} 
\par
The experiment is shown in fig. \ref{fig:setup}.  An argon-ion 
laser ($351.1\,$nm)
is used to pump a $2\,$mm thick BBO crystal.  
The crystal is
cut for type-II phase matching at 
$\lambda_{1}=\lambda_{2}=702.2\,$nm.  
To generate a first-order HG mode, a $25\,\mu$m diameter wire was placed inside the laser cavity.  The wire breaks the cylindrical symmetry of the laser cavity, forcing the laser to operate in an HG mode with a nodal line at the position of the wire \cite{beijersbergen93}.  With the wire placed vertically, we were able to generate the mode $\mathrm{HG}_{01}$ with an output power of $\sim 30$\,mW.  The
down-converted beams leave the crystal at angles of $\sim 3^{\circ}$ 
and pass through a
half-wave plate followed by a 
$1\,$mm thick BBO compensating crystal, as described in 
\cite{kwiat95}.  A UV mirror is used to reflect the pump
beam.  A half-wave plate (HWP) and quarter-wave plate (QWP) are used 
to select between the four Bell states \cite{kwiat95}.  Trombone mirror assemblies 
are used to direct the down-converted beams onto 
a $50$-$50$ beam splitter (BS).  The relative path length is adjusted by 
moving mirror assembly $M_{1}$ with a motorized translation stage.  The BS is mounted on a translation stage and can be moved in and out of the 
down-converted beams.  This allows for polarization analysis of the 
Bell-states without the BS.  The polarization analyzers (not shown) 
are rotatable half-wave plates followed by polarizing beam splitters.  
Detectors $D_{1}$
and $D_{2}$ are avalanche photodiodes equiped with $1\,$nm FWHM
bandtwidth interference filters and $3$\,mm diameter collection 
apertures. 
Coincidence and single counts were registered 
by a personal computer.   
\par
\begin{figure}
 \includegraphics[width=5cm]{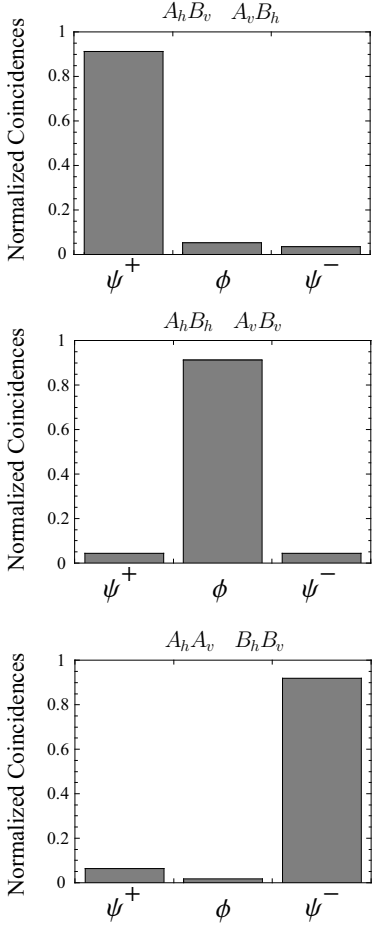}%
 \caption{ Experimental results for incomplete Bell- 
state measurement (BSM) in the coincidence basis for the three input states $\ket{\psi^{+}}$, 
$\ket{\psi^{-}}$ and $\ket{\phi}$.}
\label{fig:result}
 \end{figure}

 Polarization analysis (BS and PBS removed) of  the four Bell states generated with a $\mathrm{HG}_{01}$ 
pump beam gave interference curves (not shown) with visibilities  $\sim 0.94-0.97$, which were comparable to results when a Gaussian pump beam 
was used.  
\begin{figure}
 \includegraphics[width=5cm]{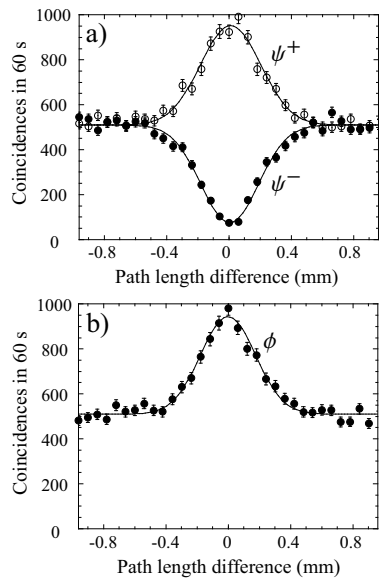}%
 \caption{ Two-photon interference for a) $\ket{\psi^{+}}$ 
($\circ$), $\ket{\psi^{-}}$ ($\bullet$) and b) $\ket{\phi}$ when the pump beam is the 
Hermite-Gaussian mode $\mathrm{HG}_{01}$. }
\label{fig:hom-psi}
 \end{figure} 
Since we are unable to distinguish between $\ket{\phi^{+}}$ and $\ket{\phi^{+}}$ with the BSA, we chose to define $\ket{\phi^{+}} \equiv \ket{\phi}$.     
Fig. \ref{fig:hom-psi} shows the HOM interference for the states 
$\ket{\psi^{\pm}}$ and $\ket{\phi}$ when the $\mathrm{HG}_{01}$ beam is used and the PBS were removed.  Visibilities of $\sim 0.85\pm0.02$ were achieved with the $\mathrm{HG}_{01}$ mode, which were slightly lower than with a Gaussian pump beam ($\sim 0.92\pm0.01$).  This was most likely due to misalignment of the wire in the laser cavity, as well as an increased sensitivity to alignment of the interferometer with the $\mathrm{HG}_{01}$ pump beam.  For the BSM, the mirror assembly was placed at position ``0".  Comparing the interference maxima and minimum at position $0$ with the constant count outside the interference region, there is a $\sim 94$\% probability that  photons in $\ket{\psi^{+}}$ and $\ket{\phi}$ ($\ket{\psi^{-}}$) end up in different (the same) outputs.     
 Fig. \ref{fig:result} shows results of the incomplete BSM for the states $\ket{\psi^{-}}$, $\ket{\psi^{+}}$ and $\ket{\phi}$.  States were discriminated with a success probability of $\sim 91$\%.     
\section{Complete Bell-state analysis in the coincidence basis}
\label{sec:cbsa}
\begin{figure}
 \includegraphics[width=6cm]{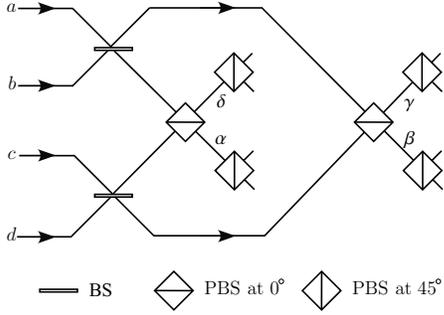}%
 \caption{Kwiat and Weinfurter's complete Bell-state 
analyzer using hyperentangled states \cite{kwiat98a}. }
\label{fig:CBSA}
 \end{figure} 
In ref. \cite{kwiat98a}, Kwiat and Weinfurter outline a scheme for 
complete Bell-state analysis using hyperentanglement, shown in fig. \ref{fig:CBSA}.  
Using hyperentangled states of the form 
\begin{align}
{\psi^{\pm}}&=\{ (a_{h}b_{v}\pm a_{v}b_{h}) + (c_{h}d_{v}\pm 
c_{v}d_{h})\}/2,  \label{eq:hpsi} \\
{\phi^{\pm}}&=\{ (a_{h}b_{h}\pm a_{v}b_{v}) + (c_{h}d_{h}\pm 
c_{v}d_{v})\}/2, \label{eq:hphi}
\end{align}   
where $a_{h} (a_{v})$ stands for a horizontally (vertically) 
polarized photon in mode $a$, the authors show that the final states 
are \begin{align}
\psi^{+}   =  & i( \alpha_{45} \delta_{45} - \alpha_{\overline{45}} 
\delta_{\overline{45}} + 
\beta_{45}\gamma_{45}-\beta_{\overline{45}}\gamma_{\overline{45}})/2, \\ 
\psi^{-}  = & 
(\alpha_{45}\gamma_{45}-\alpha_{\overline{45}}\gamma_{\overline{45}} + 
\beta_{45}\delta_{45}-\beta_{\overline{45}}\delta_{\overline{45}})/2, \\ 
\phi^{-}  = & - 
i(\alpha_{45}\alpha_{\overline{45}}-\beta_{{45}}\beta_{\overline{45}} + 
\gamma_{45}\gamma_{\overline{45}}-\delta_{{45}}\delta_{\overline{45}})/2, \\
\phi^{+}  =   & i(\alpha_{45}\alpha_{45} + 
\alpha_{\overline{45}}\alpha_{\overline{45}} +
\beta_{45}\beta_{45} + \beta_{\overline{45}}\beta_{\overline{45}}  + \gamma_{45}\gamma_{45} \nonumber \\ & + \gamma_{\overline{45}}\gamma_{\overline{45}}+
\delta_{45}\delta_{45} + \delta_{\overline{45}}\delta_{\overline{45}}
)/(2\sqrt{2}),
\end{align}   
where $\overline{45}$ is polarization orthogonal to $45$. Each of the four states above gives a different signature of 
detectors firing, however, a detector capable of distinguishing 
between one and two photons is required to detect $\phi^{+}$.  
\par
We wish to show how the use of an antisymmetric pump beam can improve on 
these results.  Using an antisymmetric pump beam such as the 
first-order Hermite Gaussian beam $HG_{01}$, it is easy to show that 
after the first set of beam splitters (BS), the states (\ref{eq:hpsi}), 
(\ref{eq:hphi}) are
\begin{align}
{\psi^{+}}&=\{ (a_{h}b_{v} +  a_{v}b_{h}) + (c_{h}d_{v} + 
c_{v}d_{h})\}/2, \\ 
{\psi^{-}}&=i\{ (a_{h}a_{v} - b_{v}b_{h}) + (c_{h}c_{v} - 
d_{v}d_{h})\}/2, \\
{\phi^{\pm}}&=\{ (a_{h}b_{h}\pm a_{v}b_{v}) + (c_{h}d_{h}\pm 
c_{v}d_{v})\}/2. 
\end{align}    
After the polarizing beam splitters (PBS) at $0^{\circ}$ and $45^{\circ}$, 
these states become
\begin{align}
\psi^{+}   =  & ( \alpha_{45} \gamma_{45} - \alpha_{\overline{45}} 
\gamma_{\overline{45}} + 
\beta_{45}\delta_{45}-\beta_{\overline{45}}\delta_{\overline{45}})/2, \\ 
\psi^{-}  = & 
i(\beta_{45}\gamma_{45}-\beta_{\overline{45}}\gamma_{\overline{45}} + 
\alpha_{45}\delta_{45}-\alpha_{\overline{45}}\delta_{\overline{45}})/2, \\ 
\phi^{-}  = &  (\alpha_{45}\beta_{\overline{45}} + 
\alpha_{\overline{45}}\beta_{{45}} +
 \gamma_{45}\delta_{\overline{45}} + \gamma_{\overline{45}}\delta_{{45}})/2, \\
\phi^{+}  = & 
(\alpha_{45}\beta_{{45}}+\alpha_{\overline{45}}\beta_{\overline{45}} + 
\gamma_{45}\delta_{{45}}+\gamma_{\overline{45}}\delta_{\overline{45}})/2. 
\end{align}    
Each of these states has its own signature of detectors firing 
\emph{in coincidence}.    
\section{Conclusion}
\label{sec:conc}
We have shown how Hong-Ou-Mandel interference at a beam splitter can be 
controlled to facilitate Bell-state analysis.  Creating 
polarization-entangled photons with spontaneous parametric 
down-conversion using an antisymmetric Hermite-Gaussian pump beam, we invert the usual interference behavior of the Bell states.  We have shown how that this simplifies the standard methods for 
incomplete Bell-state analysis of down-converted photon pairs, removing the necessity for detectors sensitive to photon number.  
In addition, we have shown that we can improve on a previous scheme 
for complete Bell-state analysis of down-converted photon pairs using 
hyperentanglement \cite{kwiat98a}, enabling complete Bell-state 
analysis in the coincidence basis.  These results illustrate the use of 
additional degrees of freedom of the two-photon state as a control parameter in quantum state engineering.  
\par
It is important to note that these results are applicable for 
down-converted pairs only, and thus are not entirely helpful for BSM's in 
quantum teleportation \cite{bouwmeester97} or entanglement 
swapping \cite{pan98,jennewein02} protocol which use photon pairs which are not created simultaneously from a common source.  However, we emphasize 
that our results are directly applicable to quantum dense coding 
schemes \cite{bennett92,mattle96}.   
   
\begin{acknowledgments}
The authors thank the Brazilian funding agencies CNPq and CAPES.
\end{acknowledgments}

\end{document}